\documentclass[a4paper,11pt]{article}
\usepackage{pos}

\newcommand{\be}{\begin{equation}}
\newcommand{\ee}{\end{equation}}
\newcommand{\bea}{\begin{eqnarray}}
\newcommand{\eea}{\end{eqnarray}}
\newcommand{\ri}{{\rm i}}

\newcommand{\Z}{\mathbb{Z}}

\newcommand{\C}{\mathbb{C}}

\newcommand {\nn}{\nonumber}
\newcommand{\vr}{{\rm v}}
\newcommand{\figsize}{0.477}

\title{Non-standard cosmic strings with amazing profiles}

\author*{Wolfgang Bietenholz}
\author{Jos\'{e} Antonio Garc\'{\i}a-Hern\'{a}ndez}
\author{\\Victor Mu\~{n}oz-Vitelly\\}

\affiliation{Instituto de Ciencias Nucleares\\
Universidad Nacional Aut\'{o}noma de M\'{e}xico \\
A.P.\ 70-543, C.P.\ 04510 Ciudad de M\'{e}xico, Mexico}

\emailAdd{wolbi@nucleares.unam.mx}
\emailAdd{antonio.garcia.lattice@gmail.com}
\emailAdd{victor.munoz@correo.nucleares.unam.mx}

\abstract{It is an unnatural feature of the Standard Model that the
difference between the baryon number $B$ and the lepton number $L$
is conserved, which implies an exact, global symmetry.
We promote it to a naturally exact, local symmetry by coupling
the difference $B-L$ to an Abelian gauge field.
Gauge anomalies are cancelled by adding right-handed neutrinos
$\nu_{R}$. They are not sterile in this case, which
forbids the Majorana term, but we arrange for a
$\nu_{R}$-mass by adding a Higgs-type 1-component complex scalar
field. Thus we arrive at a well-motivated, modest
extension of the Standard Model (SM).

We investigate the field equations in the corresponding
gauge-Higgs sector, involving both Higgs fields and the non-standard
U(1) gauge field. This leads to a set of coupled, non-linear
differential equations, which we solve numerically. We use
cylindrical coordinates, thus focusing on
the structures of cosmic strings. For a large variety of parameters,
we identify the string profiles and energy densities.
These profiles depend on the winding number of each of the Higgs
fields. For our parameters, the characteristic string radii are
below $10^{-3}~{\rm fm}$, and the string tension is not ruled out
by observations of the Cosmic Microwave Background or gravitational
waves. As an amazing peculiarity, we discover --- for high winding
numbers --- ``overshooting'' and ``co-axial'' profile functions. In
the latter case, the profile of the standard Higgs field changes
its sign near the core of the cosmic string.}

\FullConference{43rd International Conference on High Energy Physics (ICHEP 2026)\\
30 July  to 5 August , 2026\\
Natal, Brazil \vspace*{3mm} \\
{\bf Acknowledgments:}
We thank Jo\~{a}o Pinto Barros and Uwe-Jens Wiese for
contributions to this \\ project at an early stage.
This work was supported by UNAM-DGAPA through PAPIIT projects \\
IG100219 and IG100322, and by the {\it Consejo Nacional de Humanidades,
Ciencias y Tecnolog\'{\i}as} (CONAHCYT), which has now been converted
into the {\it Secretar\'{\i}a de Ciencia, Humanidades, \\
  Tecnolog\'{\i}a e Innovaci\'{o}n} (SECIHTI).}

\begin{document}
\maketitle

\section{Prototypes of cosmic strings}

Even if a field theory does not have topological sectors, it may
still have local topological defects, as observed {\it e.g.}\ in
type II superconductors and superfluid $^{4}$He. In theory, an example
are the vortices in the 2d XY model, which pile up in $d=3$ to form
closed, global cosmic strings.

\begin{figure}[h!]
\begin{center}
\includegraphics[angle=0,width=\figsize\linewidth]{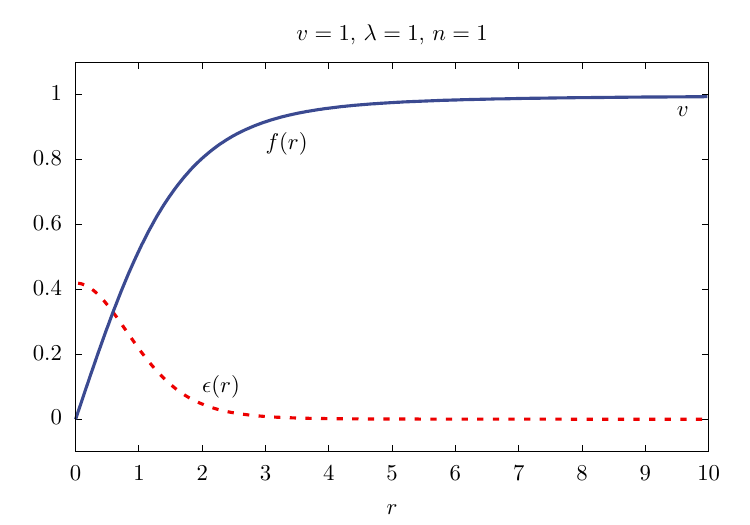}
\includegraphics[angle=0,width=\figsize\linewidth]{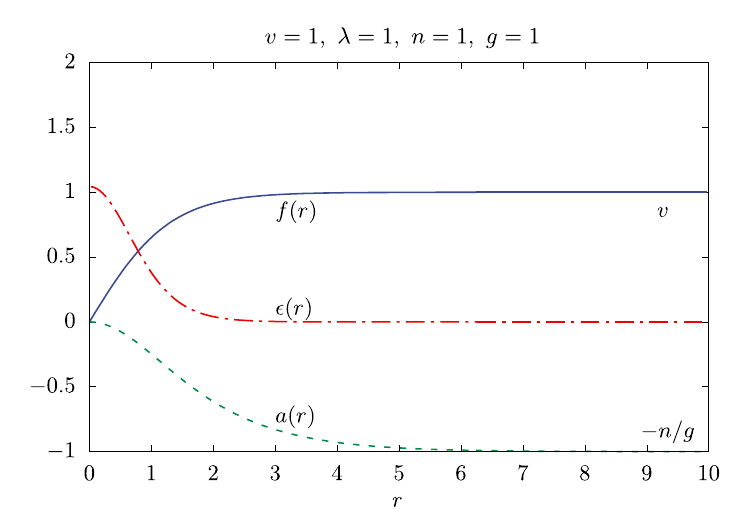}
\hspace*{-3mm}
\end{center}
\vspace*{-6mm}
\caption{Numerical solutions for a global (left) and a local (right)
  cosmic string, with $\vr =1$, $\lambda_{0}=1$, $n=1$, $g=1$: $f(r)$ and
$a(r)$ are the profile functions, and $\epsilon (r)$ is the energy density.}
\vspace*{-1mm}
\label{proto}
\end{figure}

For illustration, let us consider a 4d charged scalar field
$\chi (x) \in \C$ with the Lagrangian
\be
{\cal L} = \tfrac{1}{2} \partial^{\mu} \chi^{*} \partial_{\mu} \chi
- V(|\chi|^{2}) \ , \quad
V(|\chi|^{2}) = \tfrac{1}{2} \mu_{0}^{2} |\chi|^{2} + \tfrac{1}{4}
\lambda_{0} |\chi|^{4} \ , \quad \mu_{0}^{2} < 0, \ \lambda_{0} > 0 \ .
\ee
As an ansatz for a static solution to the field equation, we write,
in cylindrical coordinates,
\be
\chi (r,\varphi, z) = f(r) e^{\ri n \varphi} \ ,
\ee
where $f(r)$ is the profile function of this global cosmic string,
and $n \in \Z$ is its winding number.
We request the asymptotic values $f(r \to \infty ) = \vr
= \sqrt{-\mu_{0}^{2} /\lambda_{0}}$ and in the core (for $n\neq 0$)
$f(0) = 0$, which avoids a phase ambiguity.
Thus the field equation takes the non-linear form
\be
f''(r) + \frac{1}{r} f'(r) = \Big( \frac{n^{2}}{r^{2}} +
\mu_{0}^{2} + \lambda_{0} f^{2}(r) \Big) f(r) \ .
\ee
We fix the energy scale by setting $\vr =1$. The plot in
Figure \ref{proto} (left) shows, as an example, the numerical
solution for $\lambda_{0} =1$, $n=1$, where
$\epsilon (r)$ is the energy density. Note that the
energy per unit length in the $z$-direction, $E/z$, diverges
in this case logarithmically in the string radius.

When we gauge the scalar field $\chi (x)$ with a U(1) gauge field
$A_{\mu} (x)$, the Lagrangian turns into
\be
{\cal L} = \tfrac{1}{2} (D^{\mu} \chi)^{*} D_{\mu} \chi
- V(|\chi|^{2}) - \frac{1}{4} F^{\mu \nu}F_{\mu \nu} \ , \quad
D^{\mu} = \partial^{\mu} - i g A^{\mu} \ ,
\ee
and we obtain the coupled, non-linear field equations
\begin{eqnarray}
(D^{\mu} D_{\mu} + \mu_{0}^{2} + \lambda_{0}|\chi|^{2}) \chi &=& 0 \ , \nn \\
D^{\mu} F_{\mu \nu} + \tfrac{\ri g}{2}
((D_{\nu} \chi)^{*} \chi - \chi^{*} D_{\nu} \chi ) &=& 0 \ .
\end{eqnarray}
We again consider a cylindrical ansatz, with $A_{0} = A_{r} =0$ and
tangential $\vec A$,
\be
\chi = f(r) \exp (\ri n \varphi ) \ , \quad
\vec A = \frac{a(r)}{r} \hat \varphi , \quad a(0) = 0 \ .
\ee
An example for a solution is shown in Figure \ref{proto}
(right). This is a local cosmic string, with $\mu_{0}^{2}=-1$,
$\lambda_{0}=1$, $n=1$, $g=1$, and with a finite ratio $E/z$.

\section{Kibble mechanism for the formation of cosmic strings}

We may assume that topological defects were omnipresent in the
very early Universe. Under adiabatic cooling they would have
disappeared, but under rapid cooling part of them could have
persisted, and they could still persist thanks to topological
stability.

More explicitly, about $10^{-12}~{\rm sec}$ after the Big Bang the
Higgs field acquired a non-zero vacuum expectation value (VEV). In
separated regions, which were causally disconnected, the VEV had
independent complex phases. Kibble referred to vortices in the
interfaces between these regions, which could have piled up to
form strings \cite{Kibble}. Such
cosmic strings have not been observed so far, but the Kibble
mechanism is experimentally confirmed --- along with
Zurek's scaling law \cite{Zurek} --- in non-linear optical systems
\cite{Ducci}, Josephson junctions of superconductors \cite{Monaco},
manganites \cite{GriffinLin} and superfluids \cite{RystiLee}.
In numerical simulations, related phenomena were found in
particular in the XY model in dimensions $d=2$ \cite{JelCug} and
$d=3$ \cite{Lat23}.

\section{Extended Standard Model with a gauge $B - L$ symmetry and
  a massive right-handed neutrino}

We introduce an Abelian gauge field ${\cal A}_{\mu}$, which couples to
the conserved charges $Y$ (weak hypercharge) and $B-L$ with the linear
combination $2 h Y + \tfrac{1}{2} h' (B-L)$.
Gauge anomaly cancellation now requires balanced left and right
chiralities, which is achieved with a right-handed neutrino
$\nu_{R}$ in each fermion generation. This condition leaves two degrees
of freedom, which correspond to the couplings $h$ and $h'$, even
if we include the graviton \cite{pap}.
Interestingly, the SO(10) Grand Unified Theory (GUT) incorporates this
scenario, with the specific ratio $h'/h = -5$ \cite{Buch91}.

Since $\nu_{R}$ is not sterile in this case, the usual Majorana mass
term breaks gauge symmetry. So we provide a $\nu_{R}$-mass with an
additional Higgs-type field $\chi \in \C$, and consider
\begin{eqnarray}
{\cal L} &=& \tfrac{1}{2} (D^{\mu} \Phi )^{\dagger} D_{\mu} \Phi
+ \tfrac{1}{2} (d^{\mu} \chi )^{*} d_{\mu} \chi - V(\Phi,\chi)
- \tfrac{1}{4} {\cal F}^{\mu \nu} {\cal F}_{\mu \nu} \nn \\
D_{\mu} &=& \partial_{\mu} + \ri h {\cal A}_{\mu} , \quad
d_{\mu} = \partial_{\mu} + \ri h' {\cal A}_{\mu} \ , \quad
{\cal F}^{\mu \nu} = \partial^{\mu} {\cal A}^{\nu} -
\partial^{\nu} {\cal A}^{\mu} \ , \nn \\
V(\Phi,\chi) &=& \tfrac{1}{2} \mu^{2} \Phi^{\dagger} \Phi
+ \tfrac{1}{4} \lambda (\Phi^{\dagger} \Phi )^2
+ \tfrac{1}{2} \mu{'}^{2} \chi^{*} \chi
+ \tfrac{1}{4} \lambda' (\chi^{*} \chi )^2
+ \tfrac{1}{2} \kappa \, \Phi^{\dagger} \Phi \chi^{*} \chi \ ,
\end{eqnarray}
where $\Phi \in \C^{2}$ is the standard Higgs field with the
weak hypercharge $Y_{\Phi} = 1/2$,
and the non-standard Higgs field $\chi$ has charge $(B-L)_{\chi} = 2$.
We do not include SM fermions or gauge fields, assuming that they
do not significantly affect the cosmic strings. The Higgs VEVs
$\vr$, $\vr'$ should make the mass $m_{\cal A} = h\vr + h' \vr'$
sufficiently large to explain that ${\cal A}_{\mu}$ has not been observed.
The existence of a ground state requires $\kappa^{2} < \lambda \lambda'$.

Here the cylindrical ansatz
$
\Phi = \left( \begin{array}{c} 0 \\ 1 \end{array} \right)
\phi(r) e^{\ri n \varphi } \ , \
\chi = \xi(r) e^{\ri n' \varphi } \ , \
\vec {\cal A} = \tfrac{a(r)}{r} \hat \varphi \
$
leads to a system of three coupled, non-linear field equations \cite{pap},
\begin{eqnarray}
&& \hspace*{-8mm} \phi ''(r) + \tfrac{1}{r} \phi'(r) =
\Big[ \tfrac{(n + ha)^{2}}{r^{2}} + \mu^{2} + \lambda \phi^{2} +
\kappa \xi^{2} \Big] \phi \ , \
\xi ''(r) + \tfrac{1}{r} \xi'(r) =
\Big[ \tfrac{(n' + h'a)^{2}}{r^{2}} + \mu{'}^{2} + \lambda' \xi^{2} +
\kappa \phi^{2} \Big] \xi \nn \\ 
&& \hspace*{-8mm}
a ''(r) - \tfrac{1}{r} a'(r) = h \phi^{2}(n + ha) + h' \xi^{2}
(n' + h'a) \ ,
\end{eqnarray}
with the core boundary conditions
\be
\phi (0) = 0 ~~ ({\rm if}~n \neq 0) , \quad
\xi (0) = 0 ~~ ({\rm if}~n' \neq 0) , \quad a(0) = 0 \ . 
\ee
We assume asymptotically constant profile functions at large $r$,
\be
\phi (r \to \infty ) = \vr = \sqrt{\tfrac{\kappa \mu{'}^{2} - \mu^{2} \lambda '}
{\lambda \lambda' - \kappa^{2}} } \ , \quad
\xi (r \to \infty ) = \vr' = \sqrt{\tfrac{\kappa \mu^{2} - \mu{'}^{2} \lambda}
{\lambda \lambda' - \kappa^{2}} } \ , \quad
a(r \to \infty ) = - \tfrac{n}{h} = - \tfrac{n'}{h'} \ .
\ee
The latter implies for the SO(10) GUT: $n'/n=-5$.

In our numerical study, we choose two values of $v$. Based on
$\vr = 246~{\rm GeV}$ they set the scale, {\it e.g.}\ for the radius $r$,
\be
\vr = 0.01 ~:~ r=1 ~{\rm corresponds~to}~\,8 \cdot 10^{-6}~{\rm fm} \ ; \quad
\vr = 0.5 ~:~ r=1 ~{\rm corresponds~to}~\,4 \cdot 10^{-4}~{\rm fm} \ .
\ee
Precise solutions, interpolating the boundary values at $r=0$ and
$r=\infty$, were found by means of the damped Newton method, and checked
with iterative relaxation and Runge-Kutta integration \cite{pap}.

\vspace*{-2mm}
\section{Results for the string profiles}
\vspace*{-1mm}

\begin{figure}[h!]
\vspace*{-4mm}
\begin{center}
\includegraphics[angle=0,width=\figsize\linewidth]{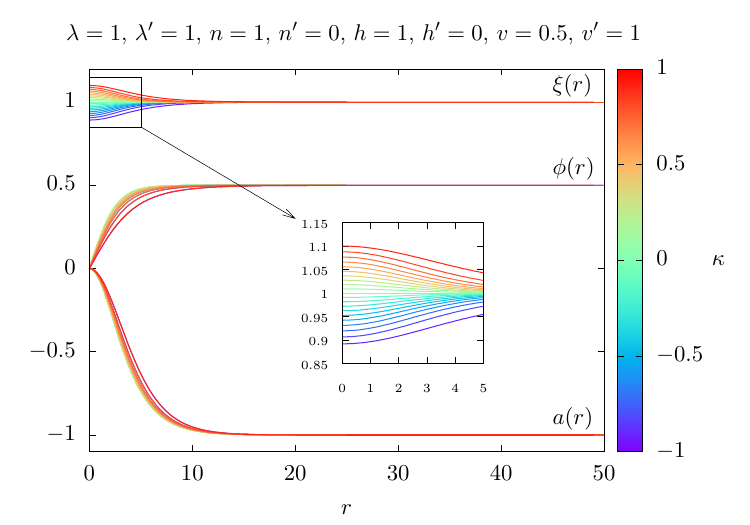}
\includegraphics[angle=0,width=\figsize\linewidth]{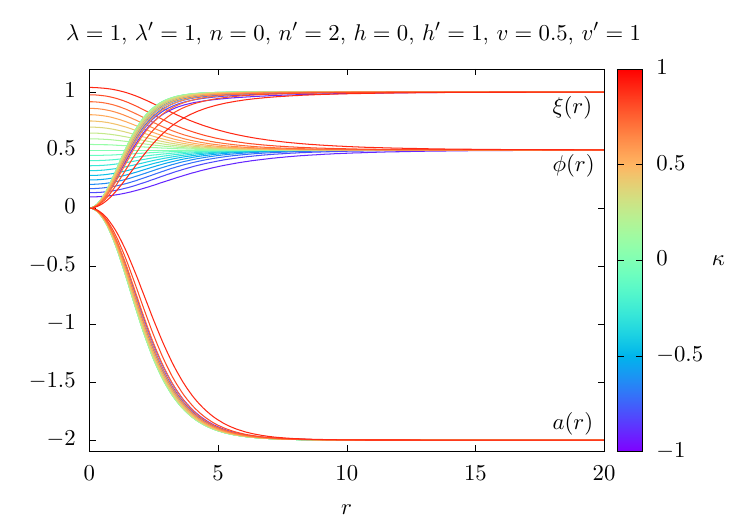}
\end{center}
\vspace*{-6mm}
\caption{We set $n'=0$ (left) and $n=0$ (right), such that
  $\xi (0) \neq 0$, but $\xi'(0) = 0$ (left), and $\phi (0) \neq 0$, but
  $\phi'(0) = 0$ (right), respectively. Here all solutions are smooth and
  monotonous. As for the energy scale, remember that
  $\phi(r \to \infty)$ corresponds to $\vr = 246~{\rm GeV}$.}
\vspace*{-1mm}
\label{nornpzero}
\end{figure}

\begin{figure}[h!]
\vspace*{-5mm}
\begin{center}
\includegraphics[angle=0,width=\figsize\linewidth]{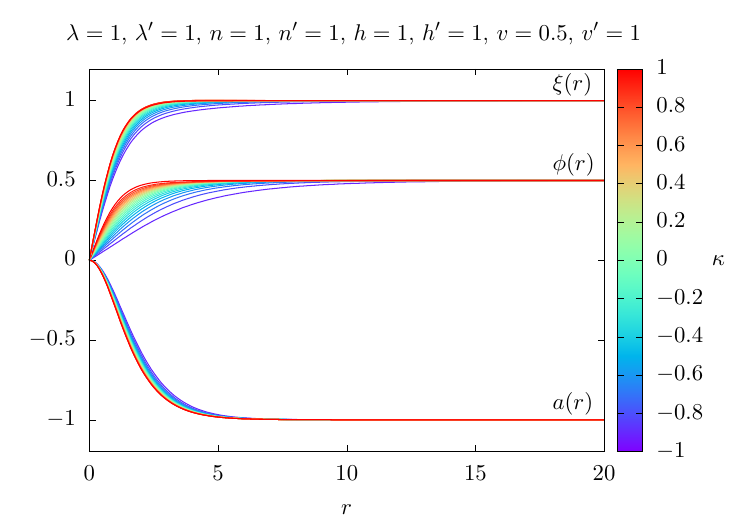}
\includegraphics[angle=0,width=\figsize\linewidth]{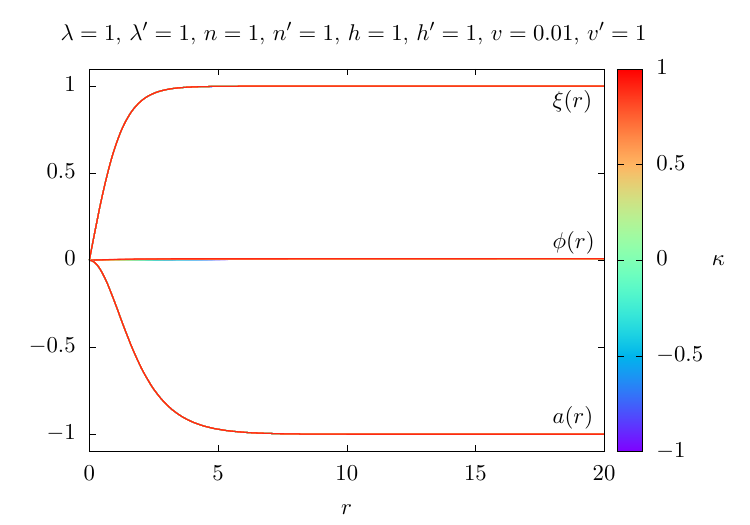}
\end{center}
\vspace*{-6mm}
\caption{We turn on windings in both Higgs fields, $n=n'=1$, at $\vr =0.5$
  (left) and $\vr = 0.01$ (right). The solutions are still smooth and
  monotonous, and the form of $\xi (r)$ is similar to the local prototype
  in Figure \ref{proto}.
  The profiles on the right represent very thin strings. They are hardly
  sensitive to $\kappa$ since $\vr'$ is very heavy, hence the $\lambda'$-term
  dominates the potential.}
\vspace*{-5mm}
\label{nnp1}
\end{figure}

\begin{figure}[h!]
\vspace*{-7mm}
\begin{center}
\includegraphics[angle=0,width=\figsize\linewidth]{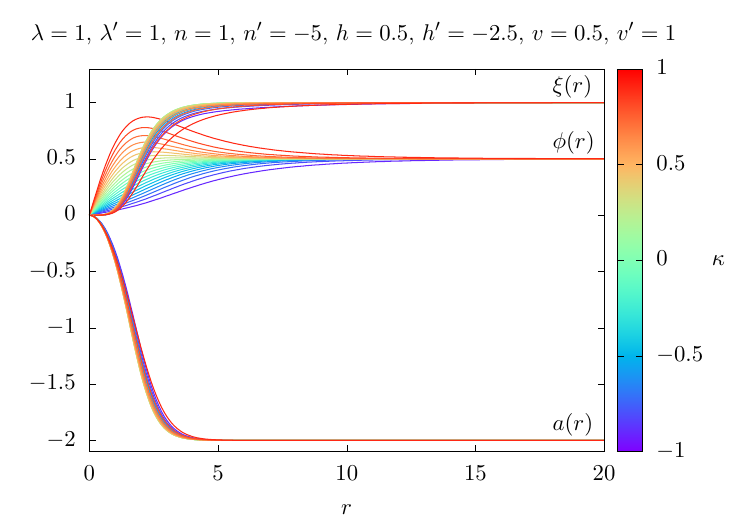}
\includegraphics[angle=0,width=\figsize\linewidth]{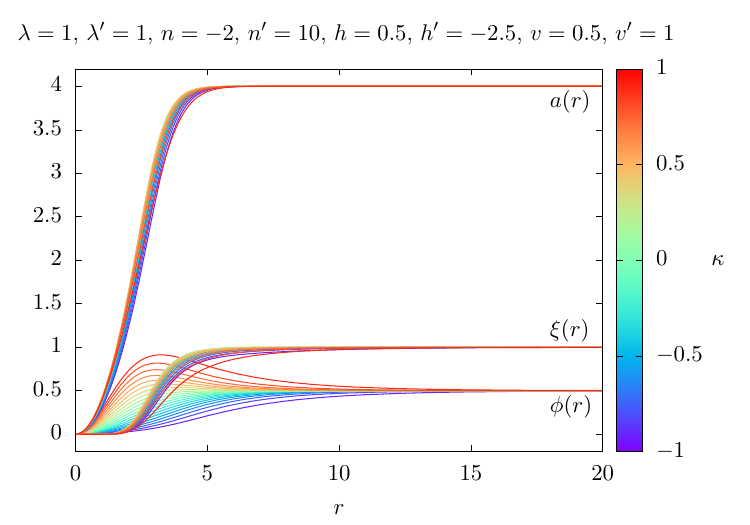}
\end{center}
\vspace*{-6mm}
\caption{We consider the SO(10) GUT scenario, which requires high
  windings: $n'=-5$ (left) and $n'=10$ (right). For $\kappa \gtrsim 0$,
  $\phi(r)$ overshoots $\vr$ near the core. In other examples,
  $\xi(r)$ can overshoot as well \cite{pap}. The effect on
  (cosmic ray) particles passing through the string is interesting.
  The gauge profile $a(r)$ flips its sign when $n'$ does
  so, which indicates that $\vec {\cal A}$ points in the direction of
  $\pm \hat \varphi$.}
\vspace*{-1mm}
\label{SO10}
\end{figure}

\begin{figure}[h!]
\vspace*{-2mm}
\begin{center}
\includegraphics[angle=0,width=\figsize\linewidth]{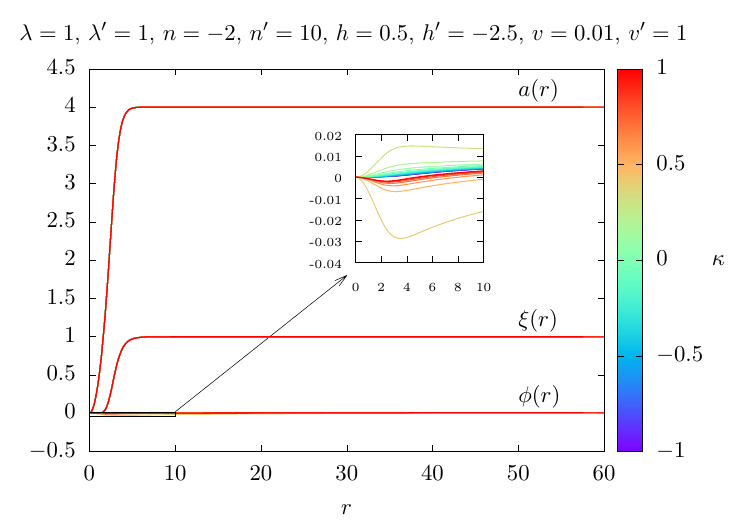}
\includegraphics[angle=0,width=\figsize\linewidth]{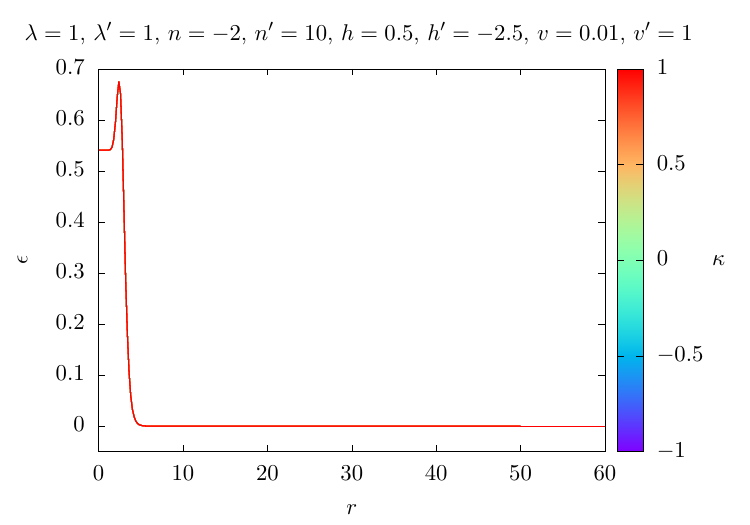}
\end{center}
\vspace*{-6mm}
\caption{We return to $\vr =0.01$ and for $\kappa \gtrsim 0.25$ we observe
  another qualitative novelty, which we call {\em co-axial}:
  $\phi (r)$ changes its sign inside the core. This effect is hardly known,
  although it was first discovered by Bogomol'nyi for unstable, standard
  cosmic strings \cite{Bogo}.
  On the right we show the radial energy density in this case:
  $\epsilon =1$ corresponds to $\approx 5 \cdot 10^{19}~ {\rm GeV/fm^{3}}$.}
\vspace*{-3mm}
\label{coaxial}
\end{figure}

For the numerical solutions of the profile functions, we scan the
allowed values of $|\kappa| < \sqrt{\lambda \lambda'}$. Examples are shown
in Figures \ref{nornpzero} to \ref{coaxial} and discussed in the
captions. The parameter sets are specified in the figure titles.

\newpage

\vspace*{-2mm}
\section{Conclusions and prospects}
\vspace*{-2mm}

We considered a modest but well-motivated SM extension:
$B-L$ is the charge coupled to a U(1) gauge field ${\cal A}_{\mu}$,
$\nu_{R}$ cancels gauge anomalies, and a non-standard Higgs field
$\chi \in \C$ provides a $\nu_{R}$-mass.
We studied the formation of cosmic strings in the gauge-Higgs sector
with $\Phi$, $\chi$, ${\cal A}_{\mu}$.
For our parameters, the characteristic radius is less than
$10^{-3}~{\rm fm}$, almost like a Nambu-Goto string.

For low winding numbers, we obtain smooth and monotonous profile
functions; higher windings may lead to overshooting and co-axial
features, including the case embedded in an SO(10) GUT.
The stability is left for future investigation.
In the latter case, the string tension amounts to
$
\mu \simeq 10^{10}~{\rm GeV}^{2} , \ G \mu \simeq 10^{-28}$
(where $G$ is the gravitation constant).
Such a string across the visible Universe would have a mass of
$10^{26}~{\rm kg} \lesssim M_{\rm Neptune}$.

Cosmic strings could have caused 1d discontinuities in the
Cosmic Microwave Background (CMB, Kaiser-Stebbins-Gott effect).
However, the search in the angular power spectrum of CMB
anisotropies by the satellites WMAP, SDSS and Planck established
the bound $G\mu < 10^{-7}$ \cite{WPW}.

Recently, gravitational waves provide a new level of precision,
at least for oscillating loops of cosmic strings:
LIGO-Virgo-KAGRA set the new bound $G\mu < 4 \cdot 10^{-15}$ \cite{LIGO3}, and
LISA expects to arrive at precision  $G\mu < {\cal O} (10^{-17})$
\cite{LISA}. Even then our scenario would not be excluded.

\end{document}